\documentclass[twocolumn,showkeys,showpacs]{revtex4-1}
\usepackage[cp1251]{inputenc}
\usepackage{amsmath}
\usepackage{amsfonts}
\usepackage{amssymb}
\usepackage{graphicx}
\usepackage{textcomp}
\usepackage{color}
\begin{document}
	
\title{Molecular dynamics simulation of graphene on Cu(111) with different Lennard-Jones parameters}
	
\author{A.V. Sidorenkov}
\author{S.V. Kolesnikov}
\email{kolesnikov@physics.msu.ru}
\author{A.M. Saletsky}
	
\affiliation{Faculty of Physics, Lomonosov Moscow State University, Moscow 119991, Russian Federation}
	
\begin{abstract}	
The interaction between graphene and copper (111) surface have been investigated using the molecular dynamics simulations. We have shown that it is possible to fit Lennard-Jones potential leading to the correct values of the binding energy and the binding distance and, at the same time, yielding experimentally observed Moir\'e superstructures. The dependencies of the binding energy, the binding distance and the graphene thickness on the parameters of the potential and the rotational angle are presented.
\end{abstract}	



\date{\today}
	
\maketitle

\section{\label{intro} Introduction}
	
Since the investigations of Novoselov et al.~\cite{Novoselov_Science,Novoselov_Nature,Novoselov_PNAS} were published, graphene has become one of the most popular subject of scientific researches due to its unique physical properties~\cite{RMP81.109,PhysRep496.109,RMP83.407,RMP83.1193,PhysRep503.77}. However, for the technical applications of graphene it is necessary to develop an effective method for producing highly crystalline wafer-scale graphene. The catalytic chemical vapor deposition (CVD) of carbon precursors is one of the most widespread methods that have been used to grow wafer-scale graphene~\cite{PhysRep542.195}. CVD is widely known to involve the decomposition of a carbon feedstock, either hydrocarbons or polymers, with the aid of heat and metal catalysts~\cite{SS603.1841,Carbon70.1}. Various metals, such as Cu, Ni, Pt, Ru, and Ir, have been proven to catalyse the growth of graphene.
	
The growth mechanism of graphene on Cu is quite different from the others, because the solubility of carbon in the Cu bulk is very low, and the mobility of carbon can be concluded to be a purely surface-based process~\cite{ACSNano6.3243}. As a result, the growth of graphene on Cu is surface mediated and the limited diffusion of carbon into the Cu bulk is occurred~\cite{NanoLett9.4268,NanoLett10.4328,UFN56.1013}. Graphene can be nucleated on various crystal facets on Cu~\cite{NanoLett11.4547}, however, the Cu(111) surface is more preferable for the synthesis of high quality monolayer graphene~\cite{ACSNano6.2319}. Two predominant graphene orientations on Cu(111) have been observed~\cite{NanoLett10.3512}: one with zero rotational angle $\Theta$ and large Moir\'e pattern ($\sim6.6$ nm periodicity) and another with $\Theta\approx7^\circ$ and smaller Moir\'e pattern ($\sim2$ nm periodicity). Moir\'e superstructures with another rotational angle ($\Theta=10.4^\circ$) are also observed~\cite{Carbon77.1082}.
	
The most widely used theoretical methods for the investigation of graphene growth on metal surfaces are the density functional theory (DFT) and the molecular dynamics (MD). DFT calculations are usually used for investigation of adsorption of carbon atoms, dimers and planar sheets of graphene on metal surfaces, while MD is more appropriate for studying graphene properties at the mesoscopic scale. A lot of recent DFT calculations~\cite{PRL99.176602,PRL101.026803,PRB81.081408R,JPCM22.485301,PRL104.186101,RSCAdv3.3046,tDMat2.014001,EPJB88.31,PRB91.045408} prove that graphene and Cu(111) surface attract each other by van der Waals forces. However, the numerical values of binding energies $E_b$ and binding distances $d$ of graphene on Cu(111) surface are quite different for different forms of the exchange-correlation functional. For example, the local density approximation (LDA) gives the following values of binging energy $E_b$ (in meV/atom): $-33$~\cite{PRL101.026803}, $-35$~\cite{PRB81.081408R}, $-39$~\cite{EPJB88.31}, $-57$~\cite{JPCM22.485301}, $-69$~\cite{tDMat2.014001}, $-70$~\cite{PRL99.176602}; the generalized gradient approximation (GGA) leads to unstable configurations with positive binding energies~\cite{tDMat2.014001,EPJB88.31}; and the usage of the van der Waals density functional (vdW-DF) approximation leads to the following $E_b$ (in meV/atom): $-38$~\cite{PRB81.081408R,tDMat2.014001}, $-180$~\cite{RSCAdv3.3046}. The values of binding distances $d$ of graphene on Cu(111) surface vary from 2.23~\r{A}~\cite{tDMat2.014001} to 3.26~\r{A}~\cite{PRL101.026803} in LDA; from 2.96~\r{A}~\cite{RSCAdv3.3046} to 3.67~\r{A}~\cite{tDMat2.014001} in vdW-DF approximation; and from 3.63~\r{A}~\cite{EPJB88.31} to 3.91~\r{A}~\cite{tDMat2.014001} in GGA. Summarizing these results, one can see that DFT calculations lead to ambiguous values of the binding energy $E_b\approx-30\div-180$ meV/atom and the binding distance $d\approx2.2\div3.9$ \r{A} of graphene on Cu(111). This ambiguity dives us some freedom in the fitting of interatomic potentials.
	
MD simulations of graphene on Cu surfaces allow to investigate such interesting phenomena as the formation of Moir\'e superlattices~\cite{Carbon77.1082,Carbon50.3055}, peeling and folding of graphene~\cite{Carbon50.3055}, interaction of graphene with metal clusters~\cite{CPL472.200}, and also jumping of metal nanodroplets~\cite{PRE83.041603}. Usually, the Lennard-Jones (L-J) potential
\begin{equation}
V_{LJ}(r_{ij})=4\epsilon\left[\left(\frac{\sigma}{r_{ij}}\right)^{12}-\left(\frac{\sigma}{r_{ij}}\right)^{6}\right]
\end{equation}
is used to describe van der Waals interaction between carbon and Cu atoms, where $r_{ij}$ is the distance between the atoms $i$ and $j$, $\epsilon$ and $\sigma$ are the L-J parameters. The investigation~\cite{Carbon50.3055} showed that parameters $\epsilon=0.0168$ eV and $\sigma=2.2$ \r{A} lead to values of $E_b$ and $d$ which agree with the results of DFT calculations~\cite{JPCM22.485301}. However, the binding energy of graphene on Cu(111) surface has only one minimum at $\Theta=0^\circ$. This result is in contradiction with the experimental data~\cite{NanoLett10.3512}. S\"ule {\it et al.} have fitted the Abell-Tersoff-like angular-dependent potential for the C-Cu interaction~\cite{Carbon77.1082} and shown that the binding energy of graphene on Cu(111) surface has local minima at the following $\Theta$: $0.0^\circ$, $2.2^\circ$, $6.7^\circ$, $8.7^\circ$, $10.4^\circ$, and $16.1^\circ$. It is necessary to underline that only three from six orientations have been observed experimentally~\cite{NanoLett10.3512,Carbon77.1082}. Besides, we have some doubts that the Abell-Tersoff potential can correctly describe van der Waals interaction. In this paper we will show that it is possible to fit L-J potential which leads to the correct values of $E_b$ and $d$ and, at the same time, gives two different Moir\'e superstructures with rotational angles $\Theta\approx0^\circ$ and $\Theta\approx7^\circ$.
	
The paper is organized as follows. In Section~\ref{method}, we present our model used for the simulations. In Section~\ref{Result}, we concentrate on the investigation of Moir\'e superstructures of graphene on Cu(111) surface at different L-J parameters of the C-Cu interaction. We conclude our paper in Section~\ref{conc}.

\section{\label{method} Computational method}

MD simulations are applied for the investigation of Moir\'e superstructures of graphene on Cu(111) surface. Carbon and copper atoms are described as classical particles interacting through interatomic potentials. Widely used Tersoff-Brenner (T-B) interatomic potential~\cite{PRB37.6991,PRB42.9458} is used to describe the carbon-carbon interaction. In the T-B potential, the total energy of the carbon-carbon interaction is expressed as
\begin{equation}
V_{C-C}=\sum_{i}\sum_{j>i}\left(V_R(r_{ij})-\bar{B}_{ij}\cdot V_A(r_{ij})\right),
\end{equation}
where $r_{ij}$ is the distance between the carbon atoms $i$ and $j$, $V_R$ and $V_A$ are the repulsive and attractive energies
\begin{equation}
V_R(r_{ij})=\frac{D_{e}}{S-1} e^{-\sqrt{2S} \beta(r_{ij}-R_{e})}f_c(r_{ij}),
\end{equation}
\begin{equation}
V_A(r_{ij})=\frac{D_{e}S}{S-1} e^{-\sqrt{2/S} \beta(r_{ij}-R_{e})}f_c(r_{ij}),
\end{equation}
$\bar{B}_{ij}$ is the many-body coupling parameter
\begin{equation}
\bar{B}_{ij}= \frac{1}{2}(B_{ij}+B_{ji}),
\end{equation}
\begin{equation}
B_{ij}= \left[ 1+\sum_{k \ne i,j }G(\theta_{ijk})f_c(r_{ik}) \right]^{- \delta}.
\end{equation}
The cutoff function $f_c(r_{ij})$ has the form
\begin{equation} \label{cutoff}
f_c(r_{ij})=
\begin{cases}
1, & $$r_{ij}<R_{1},$$  \\
$$\frac{1}{2} \left\{ 1+ \cos{\left[ \frac{\pi (r_{ij}-R_{1})}{(R_{2}-R_{1})} \right]}  \right\}, $$ & $$R_{1}<r_{ij}<R_{2},$$ \\
0, & $$r_{ij}>R_{2},$$ \\
\end{cases}
\end{equation}
and the angle function $G(\theta_{ijk})$ is
\begin{equation}
G(\theta_{ijk})=a_0 \left( 1+ \frac{c_0^2}{d_0^2}-\frac{c_0^2}{d_0^2+[1+\cos{\theta_{ijk}}]^2} \right),
\end{equation}
where $\theta_{ijk}$ is the angle between the bonds $i-j$ and $i-k$.
We use the following parameters~\cite{PRB42.9458}: $D_{e}=6.325$ eV, $S=1.29$, $\beta=1.5$ \r{A}$^{-1}$, $R_{e}=1.315$ \r{A}, $R_{1}=1.7$ \r{A}, $R_{2}=2.0$ \r{A}, $\delta=0.80469$, $a_0=0.011304$, $c_0=19.0$, $d_0=2.5$.

The copper-copper interatomic potential is formulated in second
moment of the tight-binding approximation~\cite{Clery}. In this approximation, the attractive term $U_B$ (band energy) contains the many-body interaction. The repulsive part $U_R$ is described by pair interactions (Born-Mayer form). The total copper-copper energy $V_{Cu-Cu}$ is the sum of the band energy and repulsive part:
\begin{equation}
V_{Cu-Cu}=\sum_{i}(U_{R}^i+U_{B}^i),
\end{equation}
\begin{equation}
U_{R}^{i}=A\sum_{j}\exp \left( -p \left( \dfrac{r_{ij}}{r_{0}}-1\right) \right)f_c(r_{ij}),
\end{equation}
\begin{equation}
U_{B}^{i}=-\xi\left(\sum_{j}\exp \left( -2q \left(  \dfrac{r_{ij}}{r_{0}}-1\right) \right)f_c(r_{ij})  \right)^{1/2},
\end{equation}
where $r_{ij}$ is the distance between the copper atoms $i$ and $j$, $\xi$ is an effective hopping integral, $p$ and $q$ describe the decay of the interaction strength with distance between atoms, and $r_0$ and $A$ are adjustable parameters of interatomic interaction. The cutoff function $f_c(r_{ij})$ is the same as in the T-B potential \eqref{cutoff}. The interatomic potential reproduces the bulk and surface properties of copper. Reliability of this potential for the copper surfaces has been demonstrated~\cite{PRB77.125437,PRB80.245412,EPJB86.399}. The following parameters are used in our calculations~\cite{Clery}: $A=0.0854$ eV, $\xi=1.2243$ eV, $p=10.939$, $q=2.2799$, $r_{0}=2.5563$ \r{A}, $R_{1}=6.5$ \r{A}, $R_{2}=7.5$ \r{A}.

In our work, we examine the ability of L-J potential to describe the interaction between graphene and Cu(111) surface.
Thus, the total energy of carbon-copper interaction is
\begin{equation} 	
V_{C-Cu}=\sum_{i}\sum_{j>i}V_{LJ}(r_{ij})f_c(r_{ij}),
\end{equation}
$f_c(r_{ij})$ is the cutoff function \eqref{cutoff} with parameters $R_{1}= 6.5$, $R_{2}=7.5$. We vary the parameter $\epsilon$ from 0.008 to 0.03 eV and $\sigma$ from 1.8 to 3.6 \r{A}.

\begin{figure}[t]
\begin{center}
\includegraphics[width=0.95\linewidth]{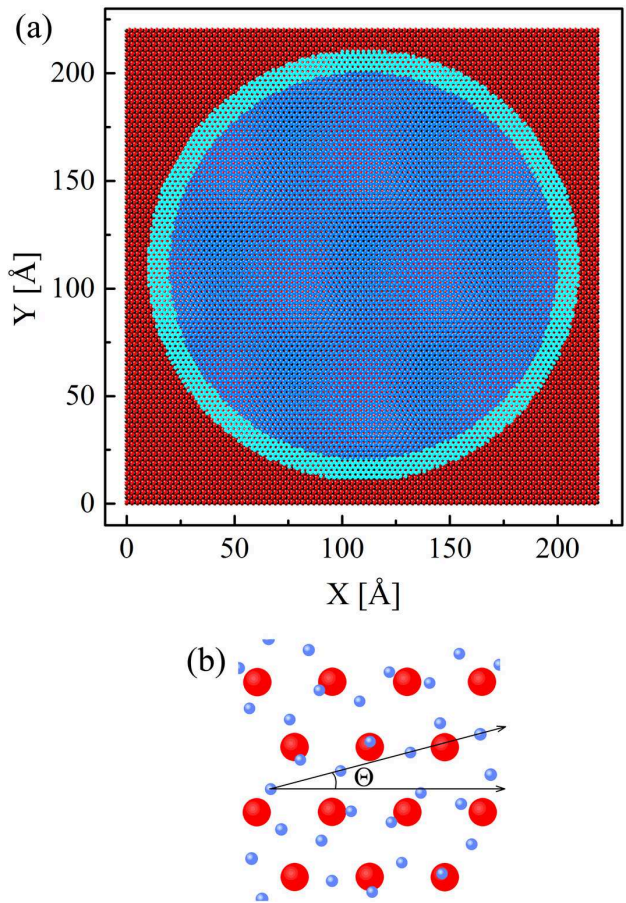}
\caption{\label{fig1} Schematic illustration of the calculation cell. (a) A graphene disc (blue color) on Cu(111) surface (red color). Carbon atoms within the edge ring (light blue color) are not accounted in the energy calculations. (b) Close view of graphene on Cu(111). Angle $\Theta$ is defined as the angle between the armchair edge of graphene and the $[\bar110]$ direction of copper (X direction of the calculation cell).
}
\end{center}
\end{figure}

The Moir\'e superstructures of graphene on Cu(111) surface are computed by means of the MD. The Cu slab consists of eight layers with 8600 atoms each. Two bottom layers are fixed and periodic boundary conditions are applied in the surface plane. A graphene disk with the diameter of 20 nm is put onto the copper surface (see Figure~\ref{fig1}a). To avoid boundary effects for graphene, the diameter of the disc is 2 nm smaller than the lateral length of the calculation cell. A series of MD simulations have been carried out for the different parameters $\epsilon$ and $\sigma$, and rotational angles $\Theta$, where $\Theta$ is the angle between the horizontal direction of the calculation cell and the zig-zag direction of graphene, as shown in Figure~\ref{fig1}b. We vary the angle $\Theta$  from  $0^\circ$ to  $11^\circ$. To obtain the binding energy between graphene and Cu(111) surface, we first relax the sample at the temperature of 300 K for 15 ps (30000 time steps). At this step we employ a chain of Nos\'e-Hoover thermostats to simulate the canonical ensemble~\cite{MolPhys52.255,PRA31.1695,MolPhys87.1117}. After that, the minimum of the energy of the carbon-copper system at the zero temperature is found by means of molecular statics method. To eliminate possible edge effects, the binding energy is calculated only for the inner area of the graphene disk with the diameter of 18 nm.
	
\section{Results and Discussions}\label{Result}

\begin{figure}[t]
\begin{center}
\includegraphics[width=0.95\linewidth]{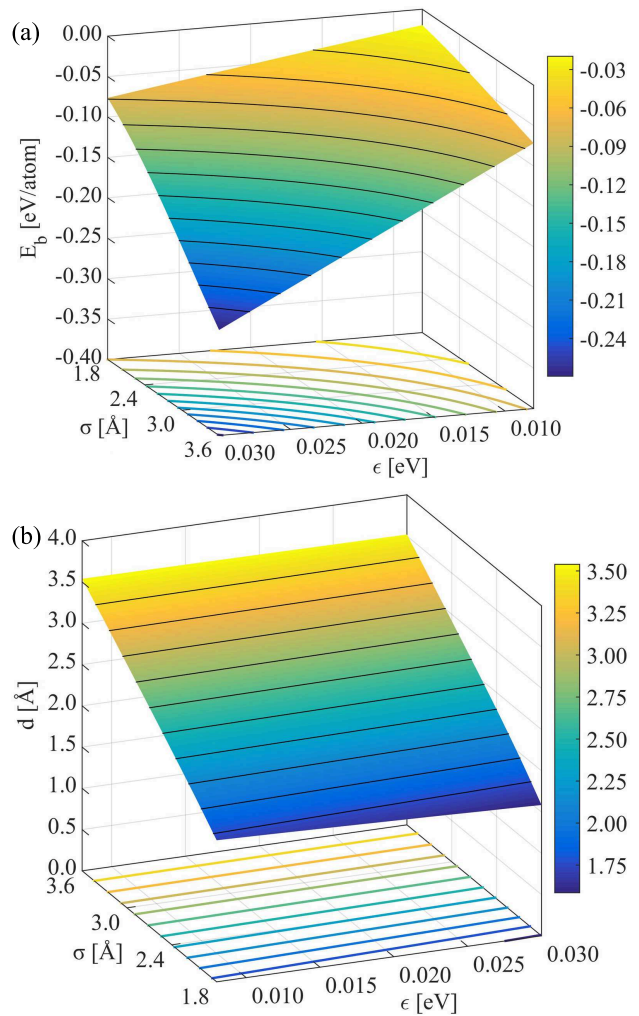}
\caption{\label{fig2} The binging energy $E_b$ normalized by the number of carbon atoms (a) and the binding distance $d$ (b) as a function of the parameters $\sigma$ and $\epsilon$ of the L-J potential. The rotational angle $\Theta=0^\circ$.
}
\end{center}
\end{figure}
	
The dependence of binding energy $E_b$ on the parameters of the L-J potential is shown in the figure~\ref{fig2}a. It is clear that $|E_b|$ increases monotonically with the growth of the parameters $\sigma$ and $\epsilon$. The dependence of $|E_b|$ on the parameter $\sigma$ becomes stronger with increasing $\epsilon$. And the dependence of $|E_b|$ on the parameter $\epsilon$ becomes stronger with increasing $\sigma$. The minimal value of the binding energy $E_b^{min}=-270$ meV/atom is reached at $\sigma=3.6$ \r{A} and $\epsilon=0.03$ eV. And the maximal value of the binding energy $E_b^{min}=-20$ meV/atom is reached at $\sigma=1.8$ \r{A} and $\epsilon=0.008$ eV. The dependence of the binding distance $d$ on the parameters of the L-J potential is shown in the figure~\ref{fig2}b. The binding distance $d$ increases in direct proportion with the parameter $\sigma$. The binding distance varies from 1.7 \r{A} ($\sigma=1.8$ \r{A}) to 3.5 \r{A} ($\sigma=3.6$ \r{A}). The dependence of $d$ on the parameter $\epsilon$ is very weak. Comparing our results with the results of DFT calculations~\cite{PRL99.176602,PRL101.026803,PRB81.081408R,JPCM22.485301,PRL104.186101,RSCAdv3.3046,tDMat2.014001,EPJB88.31} we see that there is a wide region of L-J parameters giving the appropriate values of the binding energy $E_b$ and the binding distance $d$ of graphene on Cu(111).

\begin{figure}[t]
\begin{center}
\includegraphics[width=0.95\linewidth]{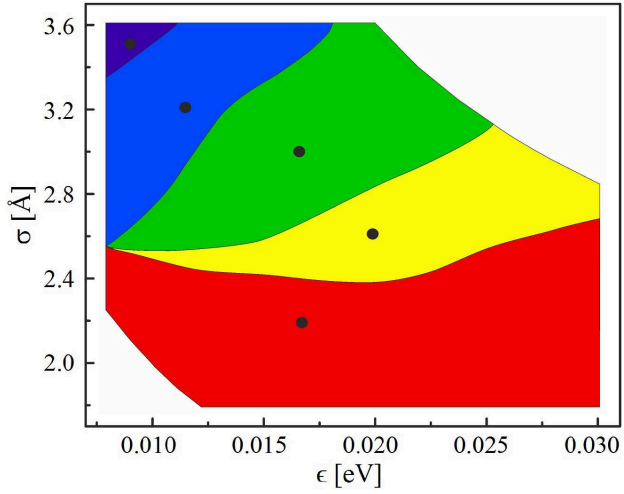}
\caption{\label{fig3} The map of the L-J parameters. The binding energy of graphene on Cu(111) has minima at different rotation angles $\Theta$ for different areas of the map. Red area: $\Theta=1^\circ$. Yellow area: $\Theta=1^\circ,8^\circ$. Green area: $\Theta=1^\circ,8^\circ,10^\circ$. Blue area: $\Theta=1^\circ,6^\circ,8^\circ,10^\circ$. Dark-blue area: $\Theta=1^\circ,3^\circ,6^\circ,8^\circ,10^\circ$. The point in the red area represents the L-J parameters used in paper~\cite{Carbon50.3055}; the points in other areas are chosen arbitrarily.
}
\end{center}
\end{figure}

\begin{figure}[t]
\begin{center}
\includegraphics[width=0.95\linewidth]{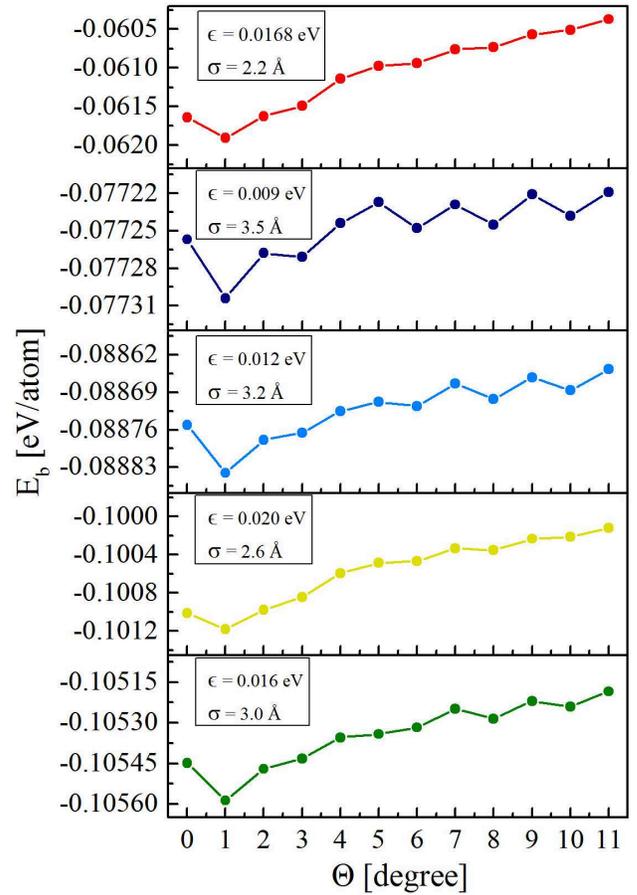}
\caption{\label{fig4} The binding energy $E_b$ normalized by the number of carbon atoms as a function of the rotation angle $\Theta$ for different L-J parameters.
}
\end{center}
\end{figure}

\begin{figure}[t]
\begin{center}
\includegraphics[width=0.95\linewidth]{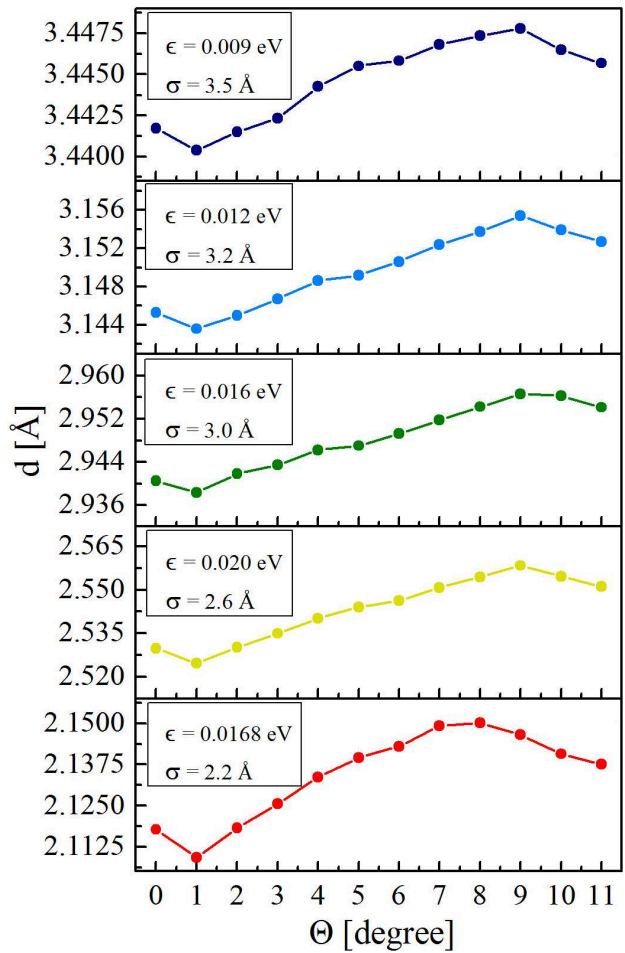}
\caption{\label{fig5} The binding distance $d$ as a function of the rotation angle $\Theta$ for different L-J parameters.
}
\end{center}
\end{figure}

\begin{figure}[t]
\begin{center}
\includegraphics[width=1.0\linewidth]{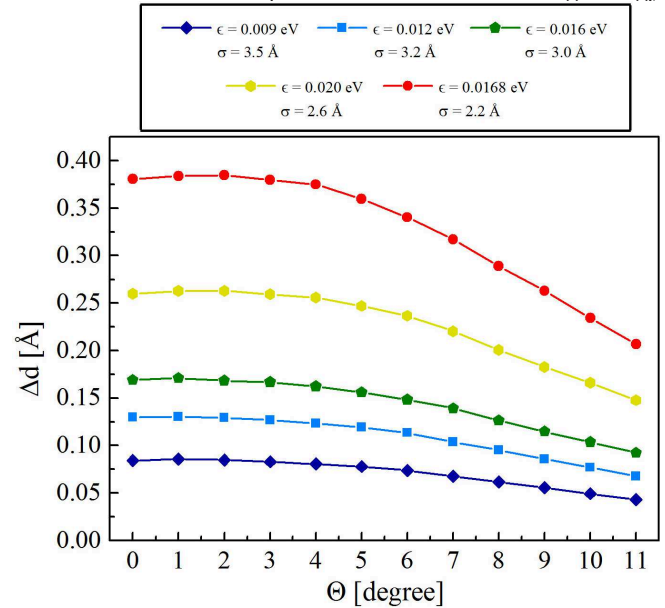}
\caption{\label{fig6} The thickness of graphene $\Delta d$ as a function of the rotation angle $\Theta$ for different L-J parameters.
}
\end{center}
\end{figure}

\begin{figure*}[t]
\begin{center}
\includegraphics[width=1.0\linewidth]{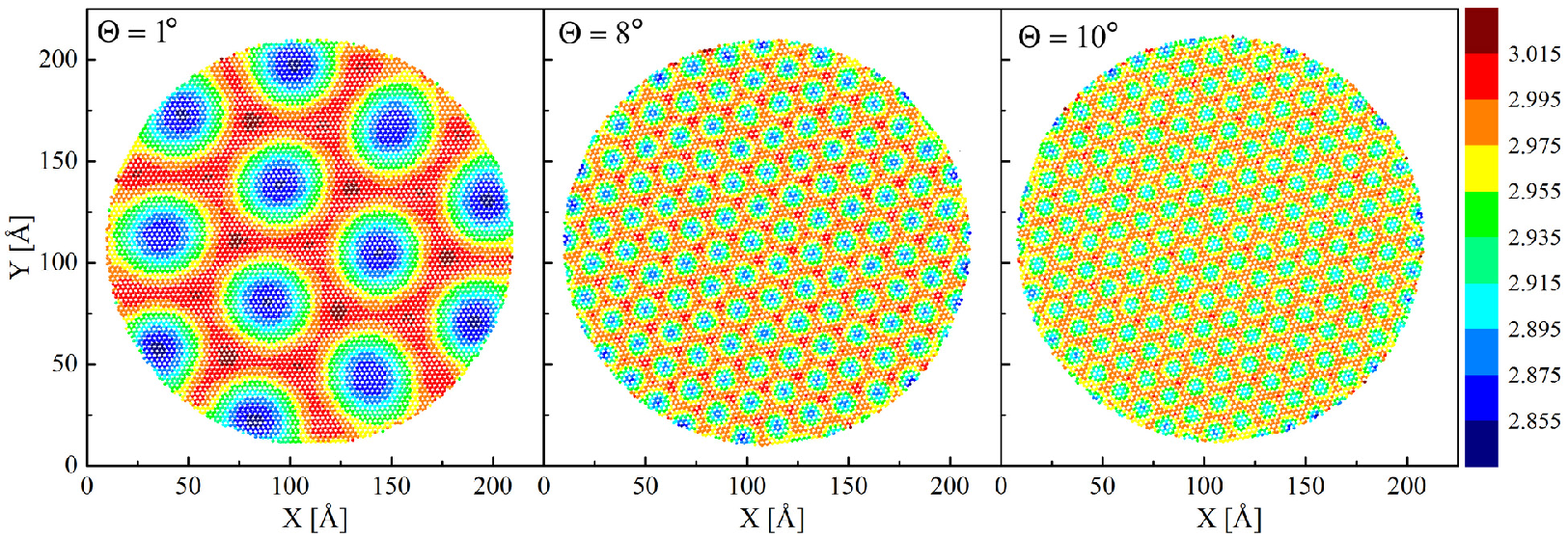}
\caption{\label{fig7} Simulated Moir\'e superstructures with different rotational angles $\Theta=1^\circ,8^\circ10^\circ$. Parameters of the L-J potential $\epsilon=0.016$ eV, $\sigma=3.0$ \r{A} lie in the green area of the map (see fig.~\ref{fig3}). The color represents the distance (in \r{A}) between the copper surface and carbon atoms.
}
\end{center}
\end{figure*}

The region of L-J parameters which gives the binding energy $E_b=-30\div-180$ meV/atom is presented in the figure~\ref{fig3} as a two-dimensional map. Different colors represent the L-J parameters leading to different local minima of the binding energy as a function of the rotation angle $\Theta$. The point in the red area represents the L-J parameters used in paper~\cite{Carbon50.3055}. The other points are chosen in the middle of each area. Figure~\ref{fig4} shows that the L-J parameters from the red area lead to only one local minimum of the binding energy $E_b=-62$ meV/atom at $\Theta=1^\circ$. This result is in a good agreement with the previous results~\cite{Carbon50.3055}: $E_b=-59$ meV/atom at $\Theta=0^\circ$. The small difference is caused by two factors: (i) we use another type of Cu-Cu interatomic potentials and (ii) we vary the rotation angle $\Theta$ with smaller step $\Delta\Theta=1^\circ$. This result is in a contradiction with the experimental data~\cite{NanoLett10.3512}. In the yellow area the binding energy has two local minima: the global minimum at $\Theta=1^\circ$ and the small local minimum at $\Theta=8^\circ$. The local minima becomes deeper with increase of the parameter $\epsilon$. The existence of two minima of the binding energy leads to two possible graphene orientations. This result is in a agreement with the  experimental data~\cite{NanoLett10.3512}. In the green area the binding energy has three local minima: the global minimum at $\Theta=1^\circ$ and two small local minima at $\Theta=8^\circ$ and $\Theta=10^\circ$. Thus, three different graphene orientations are possible. This result is in a good agreement with the experimental data~\cite{Carbon77.1082}. In the blue area the binding energy has four local minima: the global minimum at $\Theta=1^\circ$ and three small local minima at $\Theta=6^\circ,8^\circ,10^\circ$. And in the dark-blue area the binding energy has five local minima: the global minimum at $\Theta=1^\circ$ and four small local minima at $\Theta=3^\circ,6^\circ,8^\circ,10^\circ$. The graphene orientations with the rotation angles $\Theta=3^\circ,6^\circ$ have not been observed experimentally. However, the L-J parameters from the dark-blue area lead to the same results as presented in the theoretical part of the paper~\cite{Carbon77.1082}.

Figures~\ref{fig5} and~\ref{fig6} show the dependencies of the binding distance $d$ and the thickness of graphene $\Delta d$ on the rotation angle $\Theta$ for different L-J parameters. The presented dependencies have some general features. The binding distance $d$ has the minimum at $\Theta=1^\circ$ and the maximum at $\Theta\approx8^\circ\div9^\circ$. The difference $d_{max}-d_{min}$ decreases monotonically with the increase of the L-J parameter $\sigma$. In the red area $d_{max}-d_{min}\approx0.04$ \r{A} and in the dark-blue area $d_{max}-d_{min}<0.01$ \r{A}. The thickness of graphene $\Delta d$ decreases monotonically with increase of the L-J parameter $\sigma$ and the rotation angle $\Theta$.

To calculate the Moir\'e superstructures we chose the point $\epsilon=0.016$ eV, $\sigma=3.0$ \r{A} from the green area of the map. Figure~\ref{fig7} shows the Moir\'e superstructures with the rotational angles $\Theta=1^\circ,8^\circ,10^\circ$. The Moir\'e pattern has a periodicity of $\sim6.0$ nm if $\Theta=1^\circ$, $\sim1.7$ nm if $\Theta=8^\circ$, and $\sim1.4$ nm if $\Theta=10^\circ$. These results are in a good agreement with experimental data~\cite{NanoLett10.3512,Carbon77.1082}. The Moir\'e patterns for the L-J parameters from the yellow area look very similar.

\section{Conclusion}\label{conc}
	
Summarizing the results discussed above we want to underline that all L-J parameters presented in the map (Figure~\ref{fig3}) lead to the binding energies $E_b=-30\div-180$ meV/atom and the binding distances  $d=1.7\div3.5$. These values of the binding energy and the binding distance are in a agreement with the results of different DFT calculations~\cite{PRL99.176602,PRL101.026803,PRB81.081408R,JPCM22.485301,PRL104.186101,RSCAdv3.3046,tDMat2.014001,EPJB88.31}.
On the other hand, we have found five different areas on the map. In each area the binding energy $E_b$ has the different number of local minima. In particular, we have found two areas where the binding energy has two minima (at the rotation angles $\Theta=1^\circ,8^\circ$) and three minima (at the rotation angles $\Theta=1^\circ,8^\circ,10^\circ$). These results correlate with the experimental observations of the Moir\'e patterns with the rotational angles $\Theta=0^\circ,7^\circ$~\cite{NanoLett10.3512} and $\Theta=0^\circ,7^\circ,10.4^\circ$~\cite{Carbon77.1082}. In this way, we demonstrate that it is possible to fit L-J potential leading to the correct values of $E_b$ and $d$ and, at the same time, yielding experimentally observed Moir\'e superstructures.

It is very important that the L-J potential can be fitted to agree with experimental data, because the applying of a such simple potential allows substantially decrease the calculation cost for large systems. We believe that results reported here will be useful for future numerical calculations.
	
\section*{Acknowledgments}
Computational resources were provided by the Research Computing Center of the Moscow State University (MSU NIVC)~\cite{NIVC}.
	

\end{document}